# Sissy That Walk: Transportation to Work by Sexual Orientation[♠]

Sonia Oreffice[♣]

Dario Sansone[♦]

**Abstract**

We analyze differences in mode of transportation to work by sexual orientation, using the American Community Survey 2008-2019. Individuals in same-sex couples are significantly less likely to drive to work than men and women in different-sex couples. This gap is particularly stark among men: on average, almost 12 percentage point (or 13%) lower likelihood of driving to work for men in same-sex couples. Individuals in same-sex couples are also more likely to use public transport, walk, or bike to work: on average, men and women are 7 and 3 percentage points more likely, respectively, to take public transportation to work than those in different-sex couples. These differences persist after controlling for demographic characteristics, partner's characteristics, location, fertility, and marital status. Additional evidence from the General Social Survey 2008-2018 suggests that these disparities by sexual orientation may be due to lesbian, gay, and bisexual individuals caring more for the environment than straight individuals.

[♠] We thank Kitt Carpenter for helpful comments. All errors are our own.
[♣] University of Exeter, HCEO, and IZA. E-mail: s.oreffice@exeter.ac.uk
[♦] University of Exeter and IZA. E-mail: d.sansone@exeter.ac.uk

# 1. Introduction

There is by now a large and rising share (5.6%) of the US population who identifies as LGBT (Jones 2021), with a large fraction of these sexual and gender minorities active in the labor market (Badgett et al. 2021). At the same time, there are increasing environmental and health concerns associated to passive modes of transportation to work, especially in the US where up to 88% of individuals drive to work (McKenzie 2015). The vast literature on transportation, demography, health, and the environment systematically reports that passive commuting to work, and driving to work in particular, is associated with higher risk of diabetes, cholesterol, blood pressure, anxiety, depression, cardiovascular diseases, and back pain (Hoehner et al. 2012; Kylstra 2014), and that "vehicles are America's biggest air quality compromisers, producing about one-third of all U.S. air pollution" (National Geographic 2019).

However, there is no study on choices of transportation to work by sexual orientation, with scant anecdotal evidence that when it comes to the environment, LGBTQ+ adults are "greener" (Steinberg 2011). Our analysis estimates whether there are any differences in mode of transportation to work by sexual orientation, using data from the American Community Survey 2008-2019. We compare work transportation choices between individuals in same-sex and different-sex couples, highlighting differences in driving to work (by auto, truck, van, or motorcycle), in taking public transportation, in walking or biking to work, and in working from home.

Previous studies of same-sex couples present both similarities to and differences from different-sex households, also in terms of labor supply and work patterns. Black et al. (2007) assumed that families' preferences do not systematically differ by sexual orientation. They instead emphasized differences in biological constraints affecting homosexuals' fertility, location, household specialization and human capital choices. Giddings et al. (2014) showed that the specialization gap between same-sex and different-sex couples narrows across birth cohorts. Oreffice (2011) showed that same-sex couples have similar labor supply responses to intra-household bargaining power to heterosexual couples. While lesbian women earn higher wages than heterosexual women, and gay men earn lower wages than heterosexual men, some (but not all) of these differences are decreasing over time (Badgett, Carpenter and Sansone, 2021). Our analysis of modes of transportation to work represents an additional piece of evidence on households' labor market outcomes by sexual orientation, while also shedding light on general transportation patterns and why individuals may drive so much.

Although there is a lack of (large) datasets containing information on modes of transportation to work, labor market outcomes and sexual orientation, same-sex couples can be identified in the American Community Survey (ACS) by matching household heads with their same-sex spouses or unmarried partners. A large body of research confirms that the vast majority of individuals in same-sex couples in the ACS are indeed sexual minorities in a romantic relationship (Badgett et al. 2021; Black et al. 2007). We exploit the availability of the variable "Means of transportation to work", reporting the primary means of transportation to work over the course of the week preceding the interview. This information is available for the respondent but also for their unmarried partner or spouse, if present and working. We build the most recent and largest sample with detailed demographic, labor, and transportation to work information on same-sex partners/spouses, along with standard samples of different-sex partners/spouses, focusing on adult individuals aged 18 to 64 and employed.

We estimate significant differences by couple type in all work transportation arrangements, among men and women, with the largest absolute difference reported for driving to work, and the smallest one for working from home. Men and women in same-sex couples are less likely to drive to work than those in different-sex couples, but more likely to take public transportation or do active commuting such as walking or biking to work. The male disparity is always larger than the female one: the average gap represents a 13% lower likelihood in driving to work for gay men (4% for lesbian women), whereas for the other less popular work transportation arrangements the average gaps correspond to a higher likelihood of 69% or more. Our additional analysis of supplementary data indeed suggests that these differences may be due to lesbian, gay, and bisexual individuals having stronger environmental preferences than straight individuals.

We hope that this study may also provide guidance to policymakers devising environmental and transportation policies aimed at reducing car usage, and health campaigns targeting various communities in the US and abroad, but also to companies targeting healthier and productive work environments and schedules.

## 2. Data and Empirical Specification

Our main dataset is the version of the ACS publicly available through IPUMS-USA (Ruggles et al. 2021). The ACS is a nationally representative repeated cross-section conducted every year since 2000, with demographic, economic, social, work and housing information. Since 2005, it has included a 1% random sample of the US population. Although the ACS does not contain direct questions on sexual orientation, it is possible to identify unmarried same-sex couples living together. Indeed, household members can be classified as "unmarried partners" when recording their relationships to the household head, because roommates and unmarried partners are treated as two separated categories. Since 2012 same-sex couples have been allowed to report their actual marital status (between 2000 and 2012, same-sex married spouses were imputed as unmarried partners).[1]

We use data until 2019, which is the latest available wave. We start from 2008 because the US Census Bureau implemented several changes between 2007 and 2008 to reduce the number of different-sex couples misclassified as same-sex, often due to reporting errors in the sex question, which resulted in more reliable estimates and identification of same-sex couples (U.S. Census 2013). Observations with imputed sex or relation to the household head are dropped from our sample to further reduce such measurement errors, following common practice in this literature (Black et al. 2007). Notwithstanding these issues, the US Census and the ACS remain the largest and most reliable data on same-sex couples (Sansone 2019).

The main empirical specification estimates a linear probability model where the dependent variable is a dummy variable corresponding to one of the four modes of transportation to work under consideration. Most of the empirical analysis examines whether and how a binary indicator for being in a same-sex couple is associated to being more likely to drive to work, use public transport, bike or walk to work, or work from home. The other main regressors are state and year fixed effects, the respondent's age, race, ethnicity, and education, their partner/spouse's characteristics, the couple's marital status and the number of own children living in the household.[2] Robust standard errors are used to correct for heteroscedasticity throughout.

---

[1] Tables B1-B2 in the Online Appendix report sample sizes by year, couple type, sex, and marital status.
[2] All variables used in the empirical analysis are described in Section A of the Online Appendix. Table B4 provides summary statistics for relevant variables.

## 3. Results

Figure 1 presents key means in the mode of transportation to work by couple type and sex. All the four work transportation arrangements exhibit significant differences by couple type, with the largest absolute difference reported for driving to work, and the smallest one for working from home. The disparity among men in same-sex and different-sex couples is always larger than the one among women, but they are all significant at the 1% level.[3]

Table 1 reports the main results from the regressions of driving to work on a binary indicator for being in a same-sex couple, separately for women (Panel A) and men (Panel B). Starting from the basic correlation in Column 1, various controls are incrementally added, from state and year fixed effects (Column 2) to the respondent's age, race, ethnicity, and education (Column 3), their partner/spouse's characteristics (Column 4), the couple's marital status and the number of own children living in the household (Column 5).

On average, between 88% and 89% of our sample drives to work, so that the almost 4 percentage point reduction in the likelihood of driving to work associated to women in same-sex couples (Column 1) represents a 4% reduction with respect to women in different-sex couples. Even more striking are the disparities among men reported in the bottom panel, which are three times as large as those among women. The estimated coefficients indicate a reduction of 12 percentage points in the likelihood of driving to work associated to men in same-sex couples, corresponding to 13% lower propensity to drive to work with respect to men in different-sex couples. These sizable gaps are associated with important environmental and health gains. In our sample of 5.2 million partnered men interviewed between 2008 and 2019, on average 4.6 million drove to work: therefore, if men in different-sex couples had driven to work on average as frequently as men in same-sex couples, a 12 percentage point decline would have been equivalent to over 0.6 million fewer men driving. If we extrapolate the same reasoning to the 2019 ACS population means, this reduction is equivalent to over 1 million women and almost 5 million men not driving to work in a year, hugely cutting $CO_2$ emissions, negative externalities, and improving health.

All these gaps are significant at the 1% level, and robust to controlling for demographic characteristics, partner's characteristics, location, fertility, and marital status (Columns 2-5), although the magnitude of the estimates decreases. The coefficient for same-sex couples remains negative and significant also when including occupation or industry fixed effects (Table B5): thus, these differences cannot simply be attributed to sorting into jobs and different choices of workplace locations of workers in same-sex couples. A battery of robustness checks (Tables B5-B11) confirms that same-sex couples exhibit a lower propensity to drive to work.[4]

---
[3] Table B3 reports more detailed average comparisons for mode of transportation to work by sex and couple type.

Table 2 illustrates these same gaps in driving to work by marriage, parenthood, and whether couples live in the city center. These point estimates reveal that the largest gaps come from married couples without children, regardless of sex, although the disparity by couple type is always larger among men, also when couples live in the city center (Black et al. 2002).

In our Online Appendix, we present differences by couple type in active commuting such as walking and biking to work (Table B12), and in using public transportation to work (Table B13). Same-sex couples are more likely to walk or bike to work than individuals in different-sex couples, with twice as large differences among men than among women, all significant at the 1% level. The same pattern emerges for taking public transportation to work, with sizable estimates. On average 4% of individuals in the US take public transportation to work, so that a difference in around 3 percentage points among women, and 7 percentage points among men, represent a 69% and 183% increase with respect to those in different-sex couples. For biking or walking to work, the average in the US population is 2%, so that the difference in 1 percentage point among women, and 3 percentage points among men, represent a 74% and 129% increase, respectively.[5] Table B14 presents the estimated differences by couple type in working from home. Men in same-sex couples are more likely to work from home than men in different-sex couples, whereas women in same-sex couples are less likely to work from home than those in different-sex couples, although the gap is smaller.

---

[4] These include controlling for family income, being a student, working in the military, clustering SE, not using weights, using logit rather than a linear probability model, restricting the regression sample to heads, or partners/spouses, main earner in the couple, to individuals whose partner/spouse works as well, to individuals younger or older than 40, or to the 2012-2019 ACS samples. In particular, disentangling our estimates by race and ethnicity in Table B8 shows the same pattern of driving to work by couple type for Whites and Blacks, whereas among Asian and Hispanics only the male transport gap is significant.

[5] Table B15 presents multinomial logit regressions of type of transport to work (driving, public, or active), confirming our main findings: individuals in same-sex couples are more likely to use public transportation or to walk or bike rather than driving to work (women almost twice as likely and men three times).

This evidence on work transportation differences by couple type highlights that they are persistent and cannot be fully explained by factors such as marital status, presence of children in the household, occupation, industry, family income, or by focusing on those living in city centers, the youth, white individuals, or primary earners. More environmentally-conscious choices of transportation to work by same-sex couples may suggest that sexual minorities care more about the environment: indeed, we find it to be the case in the most recent General Social Survey (GSS) waves 2008-2018. Gay, lesbian and bisexual individuals can be identified from survey questions on sexual identity, while a few survey questions about the environment were asked in multiple waves, such as how interested the respondent was in environmental pollution, and whether they were concerned that the government spent too little on the environment, or on alternative energy sources. The GSS sample size is notoriously quite small for non-heterosexuals, yielding about 450 non-heterosexual individuals across all its biannual waves (Tables C1). Nevertheless, Figure 2 shows that there are significant differences by sexual orientation in terms of attitudes toward the environment: lesbian, gay, and bisexual individuals hold more environmentally-friendly views and are more concerned about the environment than straight individuals.[6]

## 4. Conclusions

These findings represent the first empirical evidence of significant and sizable differences in work traveling patterns by sexual orientation. The fact that workers in same-sex couples are less likely to drive to work and more likely to use public transportation has direct health and environmental implications that go beyond the work choices made by these households. Additional analysis suggests that these differences may be due to lesbian, gay, and bisexual individuals caring more for the environment.

We acknowledge our study's limitations inherent to the ACS data: only LGBTQ+ individuals in same-sex partnerships or marriages can be identified: un-partnered LGBTQ+ individuals are untraceable. Sexual identity is instead available in the GSS, albeit associated to a tiny sample size. In addition, the lack of data on gender identity prevents us from analyzing mode of transportation differences between transgender and cisgender individuals.

---

[6] Section A.2 in the Online Appendix provides more details on the GSS data and the variables used in this analysis. Table C2 reports more detailed average comparisons for environmental preferences by sexual orientation.

**Figure 1: Means of transportation to work by sex and couple type.**

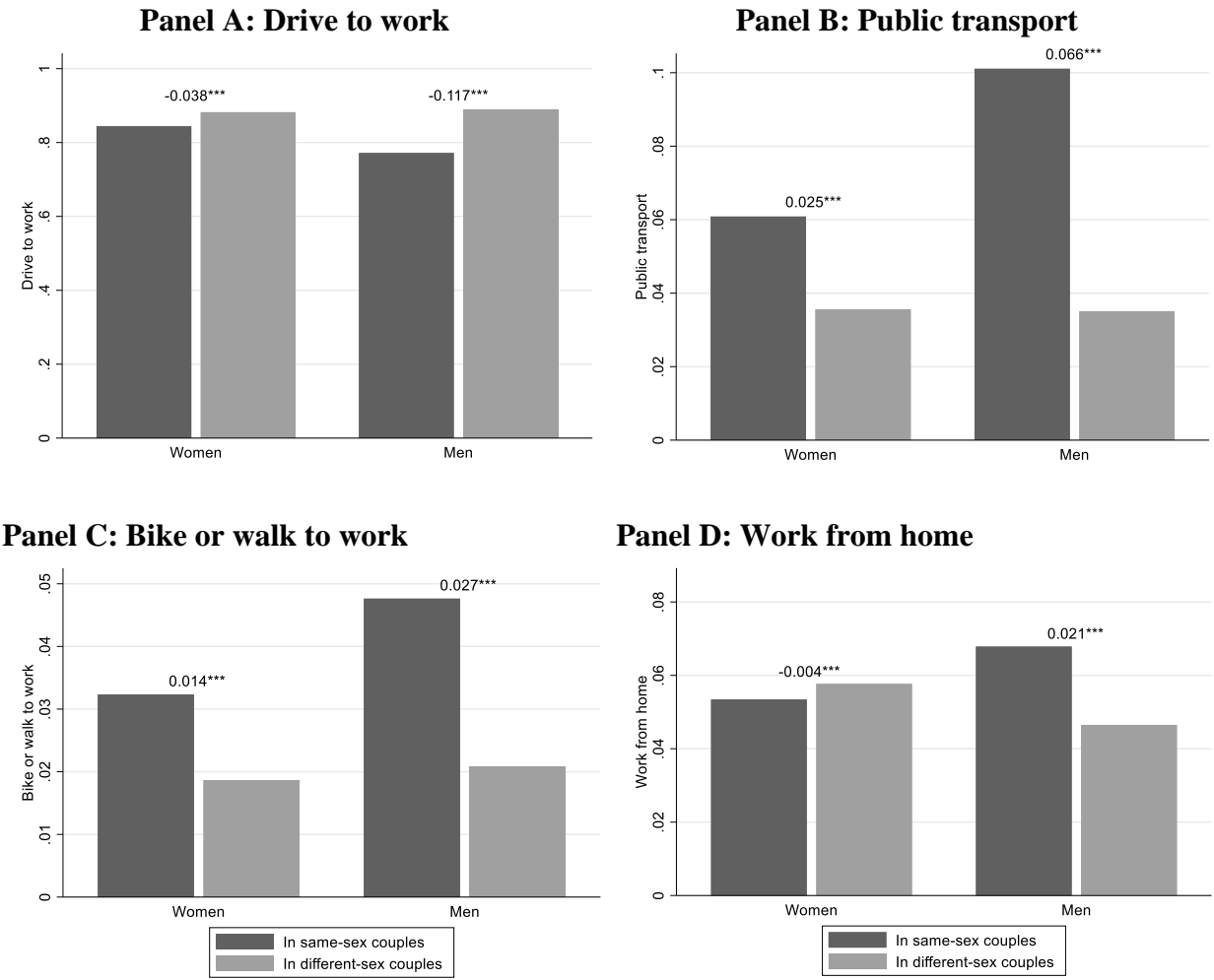

The number above each bar is the gap between the share of men or women in same-sex couples vs. in different-sex couples by mean of transportation. Weighted statistics. Source: ACS 2008-2019. * $p < 0.10$, ** $p < 0.05$, *** $p < 0.01$.

**Figure 2: Environmental preferences by sexual orientation.**

**Panel A: Too little spent on environment**

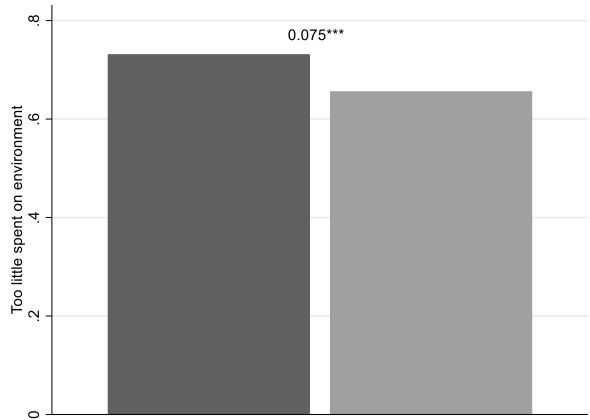

**Panel B: Too little spent on green energy**

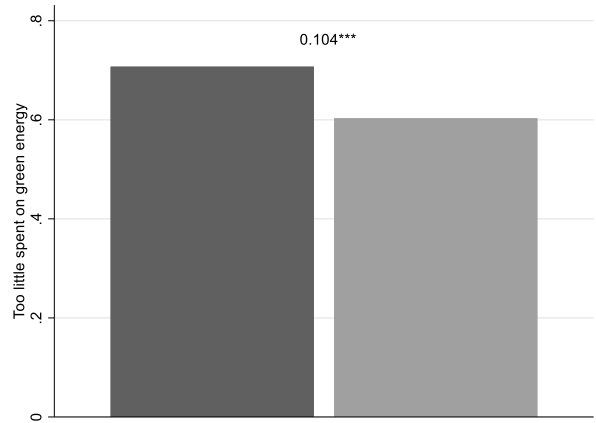

**Panel C: Very interested in environmental pollution**

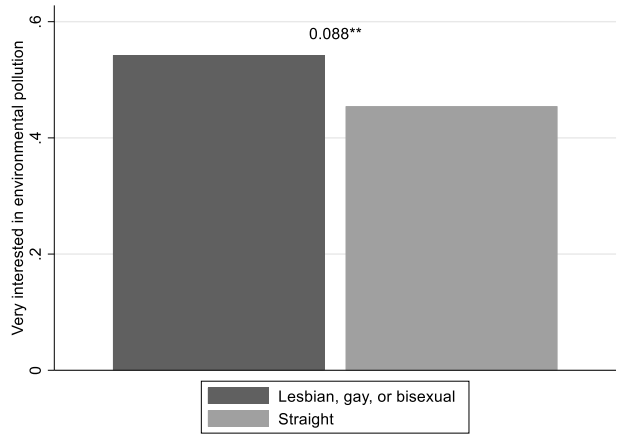

The number above each bar is the difference by sexual orientation of the share of respondents who have a certain environmental preference. Weighted statistics. Source: GSS 2008-2018. * p < 0.10, ** p < 0.05, *** p < 0.01.

**Table 1: Men and women in same-sex couples are less likely to drive to work.**

|  | (1) | (2) | (3) | (4) | (5) |
|---|---|---|---|---|---|
| *Panel A: Women in SSC and DSC* | | | | | |
| In a same-sex couple | -0.038*** | -0.030*** | -0.030*** | -0.027*** | -0.020*** |
|  | (0.002) | (0.002) | (0.002) | (0.002) | (0.002) |
| Observations | 4,411,409 | 4,411,409 | 4,411,409 | 4,411,409 | 4,411,409 |
| Mean of dependent variable | 0.881 | 0.881 | 0.881 | 0.881 | 0.881 |
| Adjusted $R^2$ | 0.000 | 0.040 | 0.044 | 0.046 | 0.047 |
| | | | | | |
| *Panel B: Men in SSC and DSC* | | | | | |
| In a same-sex couple | -0.117*** | -0.099*** | -0.091*** | -0.091*** | -0.072*** |
|  | (0.002) | (0.002) | (0.002) | (0.002) | (0.002) |
| Observations | 5,210,836 | 5,210,836 | 5,210,836 | 5,210,836 | 5,210,836 |
| Mean of dependent variable | 0.887 | 0.887 | 0.887 | 0.887 | 0.887 |
| Adjusted $R^2$ | 0.002 | 0.041 | 0.050 | 0.051 | 0.052 |
| | | | | | |
| *Controls for:* | | | | | |
| State and year FE | | ✓ | ✓ | ✓ | ✓ |
| Demographic controls | | | ✓ | ✓ | ✓ |
| Partner/spouse controls | | | | ✓ | ✓ |
| Fertility and marital status | | | | | ✓ |

"SSC" indicates same-sex couples, "DSC" indicates different-sex couples. Heteroskedasticity-robust standard errors in parentheses. Weighted regressions and statistics. Respondents younger than 18 or older than 64 have been excluded. *Demographic controls* include respondent's age, race, ethnicity, and education. *Partner/spouse controls* include spouse's or unmarried partner's age, race, ethnicity, and education. *Fertility* includes the number of own children (of any age or marital status) residing with the respondent, as well as the number of own children age 4 and under residing with the respondent. All variables are described in detail in Section A of the Online Appendix. Source: ACS 2008-2019. * $p < 0.10$, ** $p < 0.05$, *** $p < 0.01$.

**Table 2: Drive to work. Sub-sample analysis.**

|  | Married w/ children | Married w/o children | Unmarried w/ children | Unmarried w/o children | City center |
|---|---|---|---|---|---|
|  | (1) | (2) | (3) | (4) | (5) |
| *Panel A: Women in SSC and DSC* | | | | | |
| In a same-sex couple | -0.024*** | -0.062*** | -0.002 | 0.010*** | 0.003 |
|  | (0.005) | (0.004) | (0.005) | (0.003) | (0.005) |
| Observations | 1,518,968 | 1,049,278 | 144,190 | 227,662 | 452,789 |
| Mean of dependent variable | 0.880 | 0.874 | 0.893 | 0.853 | 0.710 |
| Adjusted $R^2$ | 0.040 | 0.051 | 0.055 | 0.096 | 0.234 |
| *Panel B: Men in SSC and DSC* | | | | | |
| In a same-sex couple | -0.049*** | -0.075*** | -0.017* | -0.059*** | -0.067*** |
|  | (0.007) | (0.004) | (0.010) | (0.003) | (0.004) |
| Observations | 1,972,381 | 1,092,622 | 166,510 | 235,897 | 547,612 |
| Mean of dependent variable | 0.891 | 0.876 | 0.897 | 0.841 | 0.739 |
| Adjusted $R^2$ | 0.049 | 0.056 | 0.048 | 0.102 | 0.216 |
| *Controls for:* | | | | | |
| State and year FE | ✓ | ✓ | ✓ | ✓ | ✓ |
| Demographic controls | ✓ | ✓ | ✓ | ✓ | ✓ |
| Partner/spouse controls | ✓ | ✓ | ✓ | ✓ | ✓ |
| Fertility and marital status | | | | | ✓ |

See also notes in Table 1. Source: ACS 2012-2019 in Columns 1-4; 2008-2019 in Column 5. * $p < 0.10$, ** $p < 0.05$, *** $p < 0.01$.

# Online Appendix A. Variable description

## A.1 ACS Variables

### A.1.1 Dependent variables: Means of transportation to work

Individuals were asked the following question:

> *How did this person usually get to work LAST WEEK? If this person usually used more than one method of transportation during the trip, mark (X) the box of the one used for most of the distance.*
>
> [ ] *Car, truck, or van*
> [ ] *Bus or trolley bus*
> [ ] *Streetcar or trolley car*
> [ ] *Subway or elevated*
> [ ] *Railroad*
> [ ] *Ferryboat*
> [ ] *Taxicab*
> [ ] *Motorcycle*
> [ ] *Bicycle*
> [ ] *Walked*
> [ ] *Worked at home -> SKIP to question 40a*
> [ ] *Other method*

From the answers to this question, a series of indicator variables has been constructed to record a person's main mean of transportation to work. The indicator *drive to work* includes the options "car, truck, or van" and "motorcycle".

The indicator *public transport* includes the options "bus or trolley bus"; "streetcar or trolley car"; "subway or elevated"; "railroad"; and "ferryboat". It is worth noting that a few changes were implemented in the 2019 ACS survey: the option "bus or trolley bus" was relabelled "bus"; the option "subway or elevated" was relabelled "subway or elevated rail"; the option "railroad" was relabelled "long-distance train or commuter train"; the option "light rail, streetcar, or trolley" replaced "streetcar or trolley car". Even if these changes may affect the exact option selected by an individual, they are highly unlikely to affect the proportion of individuals reporting public transport as their main mean of transportation, which is the dependent variable used in the empirical analysis.

This question on means of transportation to work was asked to all individuals age 16 or more who worked in the week preceding the interview. The aforementioned indicators have been coded as missing for individuals not working in the week preceding the interview.

### A.1.2 Key independent variable: In a same-sex couple

The ACS does not directly ask individuals about their sexual orientation. However, the ACS identifies a primary reference person, defined as "the person living or staying here in whose name this house or apartment is owned, being bought, or rented". The ACS also collects information on the relationship to the primary reference person for all members of the household, and the range of possible relationships includes husband, wife, and unmarried partner (as a different category than roommate or other nonrelative). By combining such information, it has been possible to create an indicator variable equal to one if an individual was in a same-sex couple; zero if an individual was in a different-sex couple. Both individuals married to a same-sex spouse and individuals living with a same-sex unmarried partner have been coded as individuals in same-sex couples.

It is worth nothing that, in order to reduce measurement error, in 2019 the ACS survey question explicitly distinguished between "opposite-sex husband/wife/spouse", "opposite-sex unmarried partner", "same-sex husband/wife/spouse", and "same-sex unmarried partner". In addition, the options for unmarried partners was moved higher in the list of potential relation categories, thus increasing its salience.

### A.1.3 Additional variables

*Sex* reports whether the person was male or female. Note that sex in the ACS is reported as a binary variable.

*Age* reports a person's age in years at the time of the interview. A similar variable has been constructed to report the age of a person's spouse or unmarried partner.

*Race* includes a series of indicator variables constructed to record a person's race: White, Black, Asian, or other races. The indicator *Asian* includes Chinese, Japanese, Other Asian or Pacific Islander. The indicator *other races* includes American Indian, Alaska Native, other race not listed, or individuals who selected two or three major races. A similar set of variables has been constructed to report the race of a person's spouse or unmarried partner.

*Hispanic* is an indicator equal to one if a person self-identified as Mexican, Puerto Rican, Cuban, or Other Hispanic; zero otherwise. A similar variable has been constructed to report the ethnicity of a person's spouse or unmarried partner.

*Higher Education* is an indicator equal to one if a person's highest degree completed was a Bachelor's degree or higher (Master's degree, Professional degree beyond a bachelor's degree, Doctoral degree); zero otherwise. A similar variable has been constructed to report the education level of a person's spouse or unmarried partner.

*Number of children* reports the number of own children (of any age or marital status) residing with each individual. This variable includes step-children and adopted children as well as biological children. This variable is coded as zero for people with no children present in the household.

*Number of children under age 5* reports the number of own children age 4 or under residing with each individual. This variable includes step-children and adopted children as well as biological children. This variable is coded as zero for people with no children under 5 present in the household.

*Married* is an indicator equal to one if a person is a member of a (same-sex or different-sex) married couple; zero otherwise.

*Student status* is an indicator equal to one if a person attended school or college in the 3 months preceding the interview; zero otherwise.

*In the army* is an indicator equal to one if a person reported being employed in the Armed forces (including "Armed forces: at work" and "Armed forces: with job but not at work"); zero otherwise.

*Employed* is an indicator equal to one if a person was working in the week preceding the interview; zero otherwise.

*In the labor force* is an indicator equal to one if a person was a part of the labor force, either working or seeking work, in the week preceding the interview; zero if a person was out of the labor force, or did not have a job, was looking for a job, but had not yet found one at the time of the interview.

*Total family income* reports the total pre-tax money income earned by one's family from all sources for the 12 months preceding the interview. Amounts are expressed in contemporary dollars, and not adjusted for inflation.

*Occupation* records a person's primary occupation using the IPUMS harmonized occupation coding based on the Census Bureau's 2010 ACS occupation classification scheme. Unemployed persons were to give their most recent occupation, if they had worked in the 5 years preceding the interview, otherwise they were classified as "Unemployed, with No Work Experience in the Last 5 Years or Earlier or Never Worked".

*Industry* reports the type of industry in which the person performed an occupation using the IPUMS harmonized industry coding based on the 1990 Census Bureau industrial classification scheme. Unemployed persons were to give their most recent occupation, if they had worked in the 5 years preceding the interview, otherwise they were classified as "N/A (not applicable)" or "Last worked 1984 or earlier".

### A.2 GSS Variables

Detailed information on all GSS variables are included in the GSS cumulative codebook (Smith et al. 2019). Appendix A of the codebook provides further details on the sample design and weighting. Appendix P discusses experimental forms, including information on Form 1 (the standard or "x" variant wording) versus Form 2 (the "y" variant wording"). Appendix Q discusses rotation and double sample designs, including information on the sub-samples known as "Ballots".

*Sexual orientation*. Individuals were asked the following question:

> *Which of the following best describes you?*
>   - *Gay, lesbian, or homosexual*

- *Bisexual*
- *Heterosexual or straight*
- *Don't know*

This question has been included in the GSS since 2008 in Ballots A, B, and C (i.e., in all sub-samples), but has been asked only in Ballots B and C in 2016 and 2018. Individuals who answered "Gay, lesbian, or homosexual" or "Bisexual" have been coded as "Lesbian", gay, or bisexual", while individuals who answered "Heterosexual or straight" have been coded as "Straight". This variable has been coded as missing for individuals who answered "Don't know", who refused to answer the question, or who were not asked the question.

*Spending on environment* combines the following two questions:

> *We are faced with many problems in this country, none of which can be solved easily or inexpensively. I'm going to name some of these problems, and for each one I'd like you to tell me whether you think we're spending too much money on it, too little money, or about the right amount. […] are we spending too much, too little, or about the right amount on….*
>
> *improving and protecting the environment?*
>
> *the environment?*
>
> - *Too little*
> - *About right*
> - *Too much*
> - *Don't know*

The two questions are included in Ballots A, B, and C (i.e., in all sub-samples), but they are mutually exclusive since they were used in different questionnaire versions (Form 1 versus Form 2), so no persons got asked both questions. From the answer to these questions, a series of indicator variables has been constructed to record a person's preference for environmental spending. These variables have been coded as missing for individuals who answered "Don't know", who refused to answer the question, or who were not asked the question.

*Spending on green energy*. Individuals were asked the following question:

> *We are faced with many problems in this country, none of which can be solved easily or inexpensively. I'm going to name some of these problems, and for each one I'd like you to tell me whether you think we're spending too much money on it, too little money, or about the right amount. […] are we spending too much, too little, or about the right amount on…*
>
> *developing alternative energy sources?*
>
> - *Too little*
> - *About right*

- *Too much*
- *Don't know*

This question has been included in the GSS since 2010 in Ballots A, B, and C (i.e., in all sub-samples). From the answer to this question, a series of indicator variables has been constructed to record a person's preference for green energy spending. These variables have been coded as missing for individuals who answered "Don't know", who refused to answer the question, or who were not asked the question.

*Interest in environmental pollution*. Individuals were asked to report their level of interest on issues about environmental pollution. ("Are you very interested, moderately interested, or not at all interested?"). This question has been included in the GSS since 2008. It has been asked in Ballots A, B, and C (i.e., in all sub-samples) in 2008, in Ballots A and C in 2010, in Ballots B and C in 2012 and 2014, and in Ballots A and B in 2016 and 2018. From the answer to this question, a series of indicator variables has been constructed to record a person's level interest on issues about environmental pollution. These variables have been coded as missing for individuals who answered "Don't know", who refused to answer the question, or who were not asked the question.

**Additional references for Online Appendix A**

Smith TW, Davern M, Freese J, Morgan SL, Son J, Schapiro B, Chatterjee A (2019) *General Social Surveys, 1972-2018: Cumulative Codebook* (Chicago, IL).

# Online Appendix B. ACS additional tables and figures

**Table B1: ACS sample sizes. Individuals age 18-64 in same-sex and different-sex couples.**

|       | Individuals in same-sex couples | | Individuals in different-sex couples | |
|-------|-------|-------|-----------|-----------|
|       | Women | Men   | Married   | Unmarried |
| 2008  | 5,453 | 5,079 | 997,747   | 96,396    |
| 2009  | 5,703 | 5,285 | 994,337   | 99,090    |
| 2010  | 5,733 | 5,340 | 977,773   | 106,248   |
| 2011  | 5,834 | 5,384 | 945,122   | 104,172   |
| 2012  | 6,080 | 5,603 | 942,970   | 106,056   |
| 2013  | 6,982 | 6,791 | 944,980   | 111,931   |
| 2014  | 7,380 | 7,110 | 929,088   | 113,035   |
| 2015  | 8,061 | 7,723 | 927,944   | 116,554   |
| 2016  | 8,036 | 8,021 | 922,524   | 116,246   |
| 2017  | 8,871 | 8,314 | 926,510   | 121,186   |
| 2018  | 9,137 | 8,975 | 922,169   | 122,709   |
| 2019  | 9,167 | 8,737 | 922,234   | 126,020   |
| Total | 86,437| 82,362| 11,353,398| 1,339,643 |

Notes: Sample includes all respondents (both primary reference person and unmarried partner or married spouse) in a same-sex or different-sex married/unmarried couple. Respondents younger than 18 or older than 64 have been excluded. Source: ACS 2008-2019.

**Table B2: ACS sample sizes. Individuals age 18-64 in married and unmarried couples.**

|      | Women in same-sex couples | | Men in same-sex couples | | Individuals in different-sex couples | |
|------|---------|-----------|---------|-----------|---------|-----------|
|      | Married | Unmarried | Married | Unmarried | Married | Unmarried |
| 2012 | 1,682   | 4,398     | 1,412   | 4,191     | 942,970 | 106,056   |
| 2013 | 2,383   | 4,599     | 2,095   | 4,696     | 944,980 | 111,931   |
| 2014 | 2,909   | 4,471     | 2,891   | 4,219     | 929,088 | 113,035   |
| 2015 | 4,047   | 4,014     | 3,588   | 4,135     | 927,944 | 116,554   |
| 2016 | 4,374   | 3,662     | 4,182   | 3,839     | 922,524 | 116,246   |
| 2017 | 5,296   | 3,575     | 4,681   | 3,633     | 926,510 | 121,186   |
| 2018 | 5,429   | 3,708     | 5,140   | 3,835     | 922,169 | 122,709   |
| 2019 | 5,453   | 3,714     | 4,958   | 3,779     | 922,234 | 126,020   |

Notes: Sample includes all respondents (both primary reference person and unmarried partner or married spouse) in a same-sex or different-sex married/unmarried couple. Respondents younger than 18 or older than 64 have been excluded. Marital status recorded in the ACS for same-sex couples only from 2012. Source: ACS 2012-2019.

**Table B3: Mean comparisons for mode of transportation to work by sex and couple type.**

|  | Women | | | Men | | |
|---|---|---|---|---|---|---|
|  | Same-sex couples | Different-sex couples | | Same-sex couples | Different-sex couples | |
| Variable | (1) | (2) | Gap | (3) | (4) | Gap |
| Work from home | 0.053 | 0.058 | -0.004*** | 0.068 | 0.046 | 0.021*** |
| Walk | 0.024 | 0.016 | 0.008*** | 0.036 | 0.015 | 0.021*** |
| Bike | 0.008 | 0.002 | 0.006*** | 0.011 | 0.005 | 0.006*** |
| Public transport | 0.061 | 0.036 | 0.025*** | 0.101 | 0.035 | 0.066*** |
| Taxi | 0.002 | 0.001 | 0.001*** | 0.003 | 0.001 | 0.002*** |
| Drive | 0.844 | 0.882 | -0.038*** | 0.771 | 0.888 | -0.117*** |
| N | 86,437 | 6,554,055 | | 82,362 | 6,138,986 | |

Weighted means. Sample size (N) refers to the total number of respondents in the relevant sub-group. Respondents younger than 18 or older than 64 have been excluded. Source: ACS 2008-2019. $^{*} p < 0.10$, $^{**} p < 0.05$, $^{***} p < 0.01$

**Table B4: Descriptive statistics by sex and couple type.**

|  | Women | | Men | |
|---|---|---|---|---|
|  | Same-sex couples | Different-sex couples | Same-sex couples | Different-sex couples |
| Variable | (1) | (2) | (3) | (4) |
| Age | 42.303 | 44.069 | 43.542 | 45.096 |
| White | 0.806 | 0.794 | 0.818 | 0.791 |
| Black | 0.095 | 0.074 | 0.064 | 0.085 |
| Asian | 0.027 | 0.066 | 0.045 | 0.055 |
| Other race | 0.072 | 0.067 | 0.073 | 0.069 |
| Hispanic | 0.126 | 0.146 | 0.148 | 0.152 |
| Bachelor's degree | 0.447 | 0.361 | 0.488 | 0.340 |
| Has child | 0.326 | 0.599 | 0.137 | 0.621 |
| Has child age 0-4 | 0.088 | 0.188 | 0.043 | 0.201 |
| Married | 0.483 | 0.876 | 0.460 | 0.870 |
| Student | 0.095 | 0.061 | 0.076 | 0.041 |
| In the army | 0.004 | 0.001 | 0.003 | 0.009 |
| Employed | 0.810 | 0.680 | 0.821 | 0.858 |
| In the labor force | 0.849 | 0.715 | 0.860 | 0.895 |
| Total family income | 74,282 | 101,340 | 101,851 | 101,117 |
| N | 86,437 | 6,554,055 | 82,362 | 6,138,986 |

Weighted means. Sample size (N) refers to the total number of respondents in the relevant sub-group. Respondents younger than 18 or older than 64 have been excluded. Source: ACS 2008-2019 (2012-2019 for marital status). All differences are statistically significant at the 1-percent level.

**Table B5: Drive to work. By sex and couple type. Additional controls.**

|                                | (1)        | (2)        | (3)        | (4)        |
|--------------------------------|------------|------------|------------|------------|
| *Panel A: Women in SSC and DSC* |            |            |            |            |
| In a same-sex couple           | -0.020***  | -0.019***  | -0.018***  | -0.021***  |
|                                | (0.002)    | (0.002)    | (0.002)    | (0.002)    |
| Observations                   | 4,411,409  | 4,411,409  | 4,411,409  | 4,411,409  |
| Mean of dependent variable     | 0.881      | 0.881      | 0.881      | 0.881      |
| Adjusted $R^2$                 | 0.047      | 0.099      | 0.098      | 0.047      |
|                                |            |            |            |            |
| *Panel B: Men in SSC and DSC*  |            |            |            |            |
| In a same-sex couple           | -0.072***  | -0.062***  | -0.062***  | -0.071***  |
|                                | (0.002)    | (0.002)    | (0.002)    | (0.002)    |
| Observations                   | 5,210,836  | 5,210,836  | 5,210,836  | 5,210,836  |
| Mean of dependent variable     | 0.887      | 0.887      | 0.887      | 0.887      |
| Adjusted $R^2$                 | 0.053      | 0.088      | 0.088      | 0.053      |
|                                |            |            |            |            |
| *Controls for:*                |            |            |            |            |
| State and year FE              | ✓          | ✓          | ✓          | ✓          |
| Demographic controls           | ✓          | ✓          | ✓          | ✓          |
| Partner/spouse controls        | ✓          | ✓          | ✓          | ✓          |
| Fertility and marital status   | ✓          | ✓          | ✓          | ✓          |
| Student and army status        | ✓          | ✓          | ✓          | ✓          |
| Occupation FE                  |            | ✓          |            |            |
| Industry FE                    |            |            | ✓          |            |
| Family income                  |            |            |            | ✓          |

See also notes in Table 1. Source: ACS 2008-2019. * $p < 0.10$, ** $p < 0.05$, *** $p < 0.01$.

**Table B6: Drive to work. By sex and couple type. Additional restrictions.**

|  | No students | No army | 2012-2019 | Cluster SE | No weights | Logit |
|---|---|---|---|---|---|---|
|  | (1) | (2) | (3) | (4) | (5) | (6) |
| *Panel A: Women in SSC and DSC* | | | | | | |
| In a same-sex couple | -0.018*** | -0.020*** | -0.024*** | -0.020*** | -0.021*** | -0.017*** |
|  | (0.002) | (0.002) | (0.002) | (0.002) | (0.001) | (0.002) |
| Observations | 4,139,712 | 4,404,566 | 2,940,098 | 4,411,409 | 4,411,409 | 4,411,409 |
| Mean of dependent variable | 0.880 | 0.881 | 0.877 | 0.881 | 0.881 | 0.881 |
| Adjusted $R^2$ | 0.046 | 0.047 | 0.048 | 0.047 | 0.039 | - |
| Pseudo $R^2$ | - | - | - | - | - | 0.055 |
| | | | | | | |
| *Panel B: Men in SSC and DSC* | | | | | | |
| In a same-sex couple | -0.073*** | -0.072*** | -0.075*** | -0.072*** | -0.072*** | -0.053*** |
|  | (0.002) | (0.002) | (0.002) | (0.002) | (0.002) | (0.002) |
| Observations | 5,007,438 | 5,154,713 | 3,467,410 | 5,210,836 | 5,210,836 | 5,210,836 |
| Mean of dependent variable | 0.887 | 0.886 | 0.884 | 0.887 | 0.887 | 0.887 |
| Adjusted $R^2$ | 0.052 | 0.053 | 0.055 | 0.052 | 0.046 | - |
| Pseudo $R^2$ | - | - | - | - | - | 0.063 |
| | | | | | | |
| *Controls for:* | | | | | | |
| State and year FE | ✓ | ✓ | ✓ | ✓ | ✓ | ✓ |
| Demographic controls | ✓ | ✓ | ✓ | ✓ | ✓ | ✓ |
| Partner/spouse controls | ✓ | ✓ | ✓ | ✓ | ✓ | ✓ |
| Fertility and marital status | ✓ | ✓ | ✓ | ✓ | ✓ | ✓ |

See also notes in Table 1. Source: ACS 2008-2019 (2012-2019 in Column 3). * $p < 0.10$, ** $p < 0.05$, *** $p < 0.01$.

**Table B7: Drive to work. By sex, couple type, and age group.**

|  | 18-40 | 41-64 | 25-64 |
|---|---|---|---|
|  | (1) | (2) | (3) |
| *Panel A: Women in SSC and DSC* | | | |
| In a same-sex couple | -0.022*** | -0.020*** | -0.018*** |
|  | (0.003) | (0.002) | (0.002) |
| Observations | 1,646,346 | 2,765,063 | 4,258,026 |
| Mean of dependent variable | 0.880 | 0.882 | 0.880 |
| Adjusted $R^2$ | 0.069 | 0.034 | 0.046 |
| | | | |
| *Panel B: Men in SSC and DSC* | | | |
| In a same-sex couple | -0.076*** | -0.071*** | -0.072*** |
|  | (0.003) | (0.002) | (0.002) |
| Observations | 1,852,711 | 3,358,125 | 5,090,197 |
| Mean of dependent variable | 0.888 | 0.886 | 0.886 |
| Adjusted $R^2$ | 0.076 | 0.041 | 0.052 |
| | | | |
| *Controls for:* | | | |
| State and year FE | ✓ | ✓ | ✓ |
| Demographic controls | ✓ | ✓ | ✓ |
| Partner/spouse controls | ✓ | ✓ | ✓ |
| Fertility and marital status | ✓ | ✓ | ✓ |

See also notes in Table 1. Source: ACS 2008-2019. * $p < 0.10$, ** $p < 0.05$, *** $p < 0.01$.

**Table B8: Drive to work. By sex, couple type, race, and ethnicity.**

|  | White | Black | Asian | Hispanic |
|---|---|---|---|---|
|  | (1) | (2) | (3) | (4) |
| *Panel A: Women in SSC and DSC* | | | | |
| In a same-sex couple | -0.021*** | -0.018*** | -0.002 | 0.0004 |
|  | (0.002) | (0.007) | (0.011) | (0.0048) |
| Observations | 3,650,782 | 268,726 | 261,460 | 469,924 |
| Mean of dependent variable | 0.890 | 0.869 | 0.822 | 0.864 |
| Adjusted $R^2$ | 0.032 | 0.117 | 0.121 | 0.097 |
| | | | | |
| *Panel B: Men in SSC and DSC* | | | | |
| In a same-sex couple | -0.073*** | -0.068*** | -0.075*** | -0.064*** |
|  | (0.002) | (0.009) | (0.010) | (0.005) |
| Observations | 4,313,957 | 310,190 | 281,674 | 649,412 |
| Mean of dependent variable | 0.892 | 0.884 | 0.837 | 0.893 |
| Adjusted $R^2$ | 0.044 | 0.088 | 0.106 | 0.096 |
| | | | | |
| *Controls for:* | | | | |
| State and year FE | ✓ | ✓ | ✓ | ✓ |
| Demographic controls | ✓ | ✓ | ✓ | ✓ |
| Partner/spouse controls | ✓ | ✓ | ✓ | ✓ |
| Fertility and marital status | ✓ | ✓ | ✓ | ✓ |

See also notes in Table 1. Demographic controls in these specifications include only age and education, not race or ethnicity. Source: ACS 2008-2019. * $p < 0.10$, ** $p < 0.05$, *** $p < 0.01$.

**Table B9: Drive to work. By sex, couple type, marital status, and fertility.**

|  | Only married | Only unmarried | With children | Without children |
|---|---|---|---|---|
|  | (1) | (2) | (3) | (4) |
| *Panel A: Women in SSC and DSC* | | | | |
| In a same-sex couple | -0.052*** | 0.007*** | -0.010*** | -0.019*** |
|  | (0.003) | (0.003) | (0.003) | (0.002) |
| Observations | 2,568,246 | 371,852 | 2,501,737 | 1,909,672 |
| Mean of dependent variable | 0.878 | 0.870 | 0.886 | 0.876 |
| Adjusted $R^2$ | 0.044 | 0.080 | 0.040 | 0.057 |
|  | | | | |
| *Panel B: Men in SSC and DSC* | | | | |
| In a same-sex couple | -0.079*** | -0.063*** | -0.035*** | -0.067*** |
|  | (0.003) | (0.003) | (0.005) | (0.002) |
| Observations | 3,065,003 | 402,407 | 3,226,899 | 1,983,937 |
| Mean of dependent variable | 0.886 | 0.866 | 0.895 | 0.873 |
| Adjusted $R^2$ | 0.051 | 0.083 | 0.047 | 0.062 |
|  | | | | |
| *Controls for:* | | | | |
| State and year FE | ✓ | ✓ | ✓ | ✓ |
| Demographic controls | ✓ | ✓ | ✓ | ✓ |
| Partner/spouse controls | ✓ | ✓ | ✓ | ✓ |
| Fertility | ✓ | ✓ | | |
| Marital status | | | ✓ | ✓ |

See also notes in Table 1. Source: ACS 2008-2019 (2012-2019 in Columns 1-2). * $p < 0.10$, ** $p < 0.05$, *** $p < 0.01$.

**Table B10: Drive to work. By sex, couple type, and position in household.**

|  | HH head | Spouse or partner | Main earner |
|---|---|---|---|
|  | (1) | (2) | (3) |
| *Panel A: Women in SSC and DSC* | | | |
| In a same-sex couple | -0.014*** | -0.024*** | -0.013*** |
|  | (0.002) | (0.002) | (0.002) |
| Observations | 1,846,540 | 2,564,869 | 1,753,489 |
| Mean of dependent variable | 0.873 | 0.888 | 0.880 |
| Adjusted $R^2$ | 0.047 | 0.047 | 0.062 |
|  | | | |
| *Panel B: Men in SSC and DSC* | | | |
| In a same-sex couple | -0.074*** | -0.071*** | -0.071*** |
|  | (0.003) | (0.003) | (0.002) |
| Observations | 3,173,588 | 2,037,248 | 4,004,896 |
| Mean of dependent variable | 0.885 | 0.890 | 0.887 |
| Adjusted $R^2$ | 0.053 | 0.052 | 0.054 |
|  | | | |
| *Controls for:* | | | |
| State and year FE | ✓ | ✓ | ✓ |
| Demographic controls | ✓ | ✓ | ✓ |
| Partner/spouse controls | ✓ | ✓ | ✓ |
| Fertility and marital status | ✓ | ✓ | ✓ |

See also notes in Table 1. Source: ACS 2008-2019. * $p < 0.10$, ** $p < 0.05$, *** $p < 0.01$.

**Table B11: Drive to work. By sex and couple type. Additional subsamples.**

|  | Only students | Only army | Both working |
|---|---|---|---|
|  | (1) | (2) | (3) |
| *Panel A: Women in SSC and DSC* | | | |
| In a same-sex couple | -0.037*** | -0.068*** | -0.020*** |
|  | (0.005) | (0.021) | (0.002) |
| Observations | 271,697 | 6,843 | 3,702,772 |
| Mean of dependent variable | 0.903 | 0.946 | 0.882 |
| Adjusted $R^2$ | 0.054 | 0.046 | 0.046 |
| | | | |
| *Panel B: Men in SSC and DSC* | | | |
| In a same-sex couple | -0.064*** | -0.046** | -0.074*** |
|  | (0.007) | (0.023) | (0.002) |
| Observations | 203,398 | 56,123 | 3,613,617 |
| Mean of dependent variable | 0.890 | 0.943 | 0.890 |
| Adjusted $R^2$ | 0.064 | 0.020 | 0.053 |
| | | | |
| *Controls for:* | | | |
| State and year FE | ✓ | ✓ | ✓ |
| Demographic controls | ✓ | ✓ | ✓ |
| Partner/spouse controls | ✓ | ✓ | ✓ |
| Fertility and marital status | ✓ | ✓ | ✓ |

See also notes in Table 1. Source: ACS 2008-2019. * $p < 0.10$, ** $p < 0.05$, *** $p < 0.01$.

**Table B12: Bike or walk to work. By sex and couple type.**

|                              | (1)        | (2)        | (3)        | (4)        | (5)        |
|------------------------------|------------|------------|------------|------------|------------|
| *Panel A: Women in SSC and DSC* |            |            |            |            |            |
| In a same-sex couple         | 0.014***   | 0.013***   | 0.012***   | 0.012***   | 0.006***   |
|                              | (0.001)    | (0.001)    | (0.001)    | (0.001)    | (0.001)    |
| Observations                 | 4,411,409  | 4,411,409  | 4,411,409  | 4,411,409  | 4,411,409  |
| Mean of dependent variable   | 0.019      | 0.019      | 0.019      | 0.019      | 0.019      |
| Adjusted $R^2$               | 0.000      | 0.008      | 0.009      | 0.009      | 0.010      |
|                              |            |            |            |            |            |
| *Panel B: Men in SSC and DSC* |            |            |            |            |            |
| In a same-sex couple         | 0.027***   | 0.023***   | 0.022***   | 0.022***   | 0.013***   |
|                              | (0.001)    | (0.001)    | (0.001)    | (0.001)    | (0.001)    |
| Observations                 | 5,210,836  | 5,210,836  | 5,210,836  | 5,210,836  | 5,210,836  |
| Mean of dependent variable   | 0.021      | 0.021      | 0.021      | 0.021      | 0.021      |
| Adjusted $R^2$               | 0.000      | 0.008      | 0.009      | 0.009      | 0.010      |
|                              |            |            |            |            |            |
| *Controls for:*              |            |            |            |            |            |
| State and year FE            |            | ✓          | ✓          | ✓          | ✓          |
| Demographic controls         |            |            | ✓          | ✓          | ✓          |
| Partner/spouse controls      |            |            |            | ✓          | ✓          |
| Fertility and marital status |            |            |            |            | ✓          |

See also notes in Table 1. Source: ACS 2008-2019. * $p < 0.10$, ** $p < 0.05$, *** $p < 0.01$.

**Table B13: Take public transport to work. By sex and couple type.**

|                              | (1)       | (2)       | (3)       | (4)       | (5)       |
|------------------------------|-----------|-----------|-----------|-----------|-----------|
| *Panel A: Women in SSC and DSC* |           |           |           |           |           |
| In a same-sex couple         | 0.025***  | 0.021***  | 0.022***  | 0.021***  | 0.011***  |
|                              | (0.001)   | (0.001)   | (0.001)   | (0.001)   | (0.001)   |
| Observations                 | 4,411,409 | 4,411,409 | 4,411,409 | 4,411,409 | 4,411,409 |
| Mean of dependent variable   | 0.036     | 0.036     | 0.036     | 0.036     | 0.036     |
| Adjusted $R^2$               | 0.000     | 0.078     | 0.091     | 0.092     | 0.095     |
|                              |           |           |           |           |           |
| *Panel B: Men in SSC and DSC* |           |           |           |           |           |
| In a same-sex couple         | 0.066***  | 0.054***  | 0.051***  | 0.052***  | 0.043***  |
|                              | (0.001)   | (0.001)   | (0.001)   | (0.001)   | (0.001)   |
| Observations                 | 5,210,836 | 5,210,836 | 5,210,836 | 5,210,836 | 5,210,836 |
| Mean of dependent variable   | 0.036     | 0.036     | 0.036     | 0.036     | 0.036     |
| Adjusted $R^2$               | 0.001     | 0.081     | 0.092     | 0.093     | 0.094     |
|                              |           |           |           |           |           |
| *Controls for:*              |           |           |           |           |           |
| State and year FE            |           | ✓         | ✓         | ✓         | ✓         |
| Demographic controls         |           |           | ✓         | ✓         | ✓         |
| Partner/spouse controls      |           |           |           | ✓         | ✓         |
| Fertility and marital status |           |           |           |           | ✓         |

See also notes in Table 1. Source: ACS 2008-2019. $^{*}\ p < 0.10$, $^{**}\ p < 0.05$, $^{***}\ p < 0.01$.

**Table B14: Work from home. By sex and couple type.**

|  | (1) | (2) | (3) | (4) | (5) |
|---|---|---|---|---|---|
| *Panel A: Women in SSC and DSC* | | | | | |
| In a same-sex couple | -0.004*** | -0.007*** | -0.008*** | -0.010*** | 0.0004 |
|  | (0.001) | (0.001) | (0.001) | (0.001) | (0.0010) |
| Observations | 4,411,409 | 4,411,409 | 4,411,409 | 4,411,409 | 4,411,409 |
| Mean of dependent variable | 0.058 | 0.058 | 0.058 | 0.058 | 0.058 |
| Adjusted $R^2$ | 0.000 | 0.004 | 0.009 | 0.011 | 0.014 |
| | | | | | |
| *Panel B: Men in SSC and DSC* | | | | | |
| In a same-sex couple | 0.021*** | 0.019*** | 0.015*** | 0.014*** | 0.015*** |
|  | (0.001) | (0.001) | (0.001) | (0.001) | (0.001) |
| Observations | 5,210,836 | 5,210,836 | 5,210,836 | 5,210,836 | 5,210,836 |
| Mean of dependent variable | 0.047 | 0.047 | 0.047 | 0.047 | 0.047 |
| Adjusted $R^2$ | 0.000 | 0.002 | 0.012 | 0.013 | 0.013 |
| | | | | | |
| *Controls for:* | | | | | |
| State and year FE | | ✓ | ✓ | ✓ | ✓ |
| Demographic controls | | | ✓ | ✓ | ✓ |
| Partner/spouse controls | | | | ✓ | ✓ |
| Fertility and marital status | | | | | ✓ |

See also notes in Table 1. Source: ACS 2008-2019. * $p < 0.10$, ** $p < 0.05$, *** $p < 0.01$.

**Table B15: Multinomial logit of type of transport to work (driving, public, or active). By sex and couple type.**

|  | (1) | (2) | (3) | (4) | (5) |
|---|---|---|---|---|---|
| *Panel A: Women in SSC and DSC* | | | | | |
| *Public transport* | | | | | |
| In a same-sex couple | 1.795*** | 1.725*** | 1.752*** | 1.713*** | 1.285*** |
|  | (0.035) | (0.036) | (0.038) | (0.037) | (0.029) |
| *Active transport* | | | | | |
| In a same-sex couple | 1.811*** | 1.745*** | 1.749*** | 1.717*** | 1.292*** |
|  | (0.048) | (0.047) | (0.047) | (0.047) | (0.037) |
| Observations | 4,129,831 | 4,129,831 | 4,129,831 | 4,129,831 | 4,129,831 |
| Pseudo $R^2$ | 0.001 | 0.124 | 0.150 | 0.153 | 0.160 |
| *Panel B: Men in SSC and DSC* | | | | | |
| *Public transport* | | | | | |
| In a same-sex couple | 3.332*** | 2.906*** | 2.682*** | 2.758*** | 2.053*** |
|  | (0.053) | (0.051) | (0.049) | (0.050) | (0.039) |
| *Active transport* | | | | | |
| In a same-sex couple | 2.628*** | 2.408*** | 2.240*** | 2.280*** | 1.453*** |
|  | (0.023) | (0.057) | (0.053) | (0.054) | (0.036) |
| Observations | 4,924,212 | 4,924,212 | 4,924,212 | 4,924,212 | 4,924,212 |
| Pseudo $R^2$ | 0.003 | 0.123 | 0.147 | 0.150 | 0.154 |
| *Controls for:* | | | | | |
| State and year FE |  | ✓ | ✓ | ✓ | ✓ |
| Demographic controls |  |  | ✓ | ✓ | ✓ |
| Partner/spouse controls |  |  |  | ✓ | ✓ |
| Fertility and marital status |  |  |  |  | ✓ |

See also notes in Table 1. Source: ACS 2008-2019. * $p < 0.10$, ** $p < 0.05$, *** $p < 0.01$. Relative risk ratios reported, with driving to work as base outcome. Active transport is defined as walking or biking to work.

# Online Appendix C. GSS additional tables and figures

**Table C1: GSS sample sizes. Individuals age 18-64 by sexual orientation.**

|      | Gay, lesbian, or homosexual | Bisexual | Heterosexual or straight | Don't know | No answer | Not applicable |
|------|---------|---------|---------|----|-----|-------|
| 2008 | 32      | 26      | 1,398   | 5  | 18  | 159   |
| 2010 | 27      | 31      | 1,436   | 7  | 14  | 133   |
| 2012 | 24      | 40      | 1,345   | 2  | 22  | 142   |
| 2014 | 38      | 60      | 1,770   | 5  | 29  | 109   |
| 2016 | 43      | 52      | 1,318   | 3  | 21  | 803   |
| 2018 | 29      | 47      | 1,020   | 4  | 22  | 680   |
| Total | 193    | 256     | 8,287   | 26 | 126 | 2,026 |

Notes: Respondents younger than 18 or older than 64 have been excluded. "Not applicable" includes respondents who were not asked the relevant question. Source: GSS 2008-2018.

**Table C2: Mean comparisons for environmental preferences by sexual orientation.**

| Variable | GLB (1) | Straight (2) | Gap |
|---|---|---|---|
| *Improving and protecting environment* | | | |
| Spending too little | 0.731 | 0.656 | 0.075*** |
| Spending about right | 0.197 | 0.257 | -0.060*** |
| Spending too much | 0.072 | 0.087 | -0.015 |
| *Developing alternative energy sources* | | | |
| Spending too little | 0.707 | 0.603 | 0.104*** |
| Spending about right | 0.208 | 0.313 | -0.105*** |
| Spending too much | 0.085 | 0.084 | 0.001 |
| *Interested in issues on environmental pollution* | | | |
| Very interested | 0.542 | 0.454 | 0.088** |
| Moderately interested | 0.382 | 0.455 | -0.073* |
| Not at all interested | 0.076 | 0.091 | -0.014 |
| N | 449 | 8,287 | |

Weighted means. "GLB" refers to individual who identifies as gay, lesbian, homosexual, or bisexual when asked about their sexual orientation. Sample size (N) refers to the total number of respondents in the relevant sub-group. Respondents younger than 18 or older than 64 have been excluded. Respondents who answered "don't know", who refused to answer, or who were not asked the relevant questions are not included in the above comparisons. Source: GSS 2008-2018. * $p < 0.10$, ** $p < 0.05$, *** $p < 0.01$